# Specific Heat of $Ca_{0.33}Na_{0.67}Fe_2As_2$

J. S. Kim[1], K. Zhao[2], C. Q. Jin[2], and G. R. Stewart[1]

[1]Physics Department, University of Florida, Gainesville, FL 32611-8440
[2]Institute of Physics, Chinese Academy of Sciences, Beijing 100190, China

**Abstract:** The specific heat of single crystal hole-doped $Ca_{0.33}Na_{0.67}Fe_2As_2$, $T_c^{onset}$ =33.7 K, was measured from 0.4 to 40 K. The discontinuity in the specific heat at $T_c$, $\Delta C$, divided by $T_c$ is 105 ± 5 mJ/molK$^2$, consistent with values found previously for hole-doped $Ba_{0.6}K_{0.4}Fe_2As_2$ and somewhat above the general trend for $\Delta C/T_c$ vs $T_c$ for the iron based superconductors established by Bud'ko, Ni and Canfield. The usefulness of measured values of $\Delta C/T_c$ as an important metric for the quality of samples is discussed.

I. Introduction

Although the first high temperature iron based superconductor (IBS), $LaFeAsO_{1-x}F_x$, was achieved [1] via electron doping, the role of hole doping to achieve the superconducting state in IBSs is well established [2]. The second structure in which superconductivity was found in the IBSs, tetragonal $ThCr_2Si_2$ ("122" structure), in fact involved [3] K (hole) doping in $Ba_{1-x}K_xFe_2As_2$. Since that discovery, in addition to K the cations Na, Rb, Cs, and La have been used [2] – primarily in doping the 122 $MFe_2As_2$ structure, with M=Ba, Sr, Ca and Eu.

Although doping $CaFe_2As_2$ with Na to cause superconductivity is not new, see refs. 4-7, earlier reports [4] were on polycrystals, $T_c$=34 K, and resulted in a value of the discontinuity in the specific heat, $\Delta C$, divided by $T_c$ at the superconducting transition of 66 mJ/molK$^2$. This value agreed with the Bud'ko, Ni, and Canfield (hereafter BNC) reported [8] general trend in $\Delta C/T_c$ with $T_c$ in the IBSs, a trend later confirmed for a larger number of IBS compounds [9] by Kim et al. However, this value of 66 mJ/molK$^2$ was significantly lower than results [9] for a similar $T_c$ (34.6 K) material ($Ba_{0.6}K_{0.4}Fe_2As_2$) that was in single crystal form, where $\Delta C/T_c \approx 100$ mJ/molK$^2$. Thus, the question arose: is there room in the range of sample quality for Na-doped $CaFe_2As_2$ (perhaps in single crystal material) for improvement in $\Delta C/T_c$? If so, then the more general question is, should the fitted line through almost 30 different IBS compounds [9] be understood as perhaps too low when considering 'best' quality samples?

The present work benefits from the result of several year's effort to optimize the Na concentration to achieve the maximum $T_c$ of approximately 34 K. Also, improved quality control of single crystals has increased the sharpness of the specific heat transition. Thus, the present work reports on the specific heat of optimized single crystals of $Ca_{1-x}Na_xFe_2As_2$, x=0.67, to address the question of its $\Delta C/T_c$ value compared to the BNC trend and to the results for single crystal $Ba_{0.6}K_{0.4}Fe_2As_2$.

## II. Experimental

$Ca_{0.33}Na_{0.67}Fe_2As_2$ single crystals were prepared by the solid-state reaction method with NaAs as flux. First, $Ca_{0.33}Na_{0.67}Fe_2As_2$ polycrystals was synthesized by using CaAs, $Na_3As$, FeAs, and high purity Fe powders as starting materials. The detailed process and conditions of synthesis for polycrystals were described in ref. 4. Then, $(Ca_{0.33}Na_{0.67})Fe_2As_2$ polycrystals and NaAs powder were mixed in the molar ratio 1: 10. The mixture was loaded into an evacuated alumina tube with 99% purity before being sealed into a Nb tube. The Nb tube was in turn sealed into an evacuated quartz tube, which is heated at 970 ºC for 20 hours and slowly decreased to 800 ºC at a rate of 2 ºC/h. Platelike crystals are obtained with typical size 3 × 2 × 0.05 $mm^3$. Single crystals were characterized by x-ray diffraction using Cu Kα radiation. Further characterization of the crystals was carried out using temperature dependent magnetic susceptibility, with samples showing a full shielding (zero-field cooled) effect.

$Ca_{1-x}Na_xFe_2As_2$ samples have been reported to be insensitive to exposure to air. However, in measuring specific heat the samples are usually attached using either Wakefield Grease or GE7031 varnish which can [10] react with IBS materials. Thus, the dc magnetic susceptibility was checked for single crystals of $Ca_{1-}$

$_x$Na$_x$Fe$_2$As$_2$ with these substances applied; no significant broadening or shift of the superconducting transition temperature with Wakefield Grease was detected, as shown in Fig. 1, while a slight (10%) increase in transition width was detected if GE7031 varnish was used. Specific heat was measured using established techniques [11].

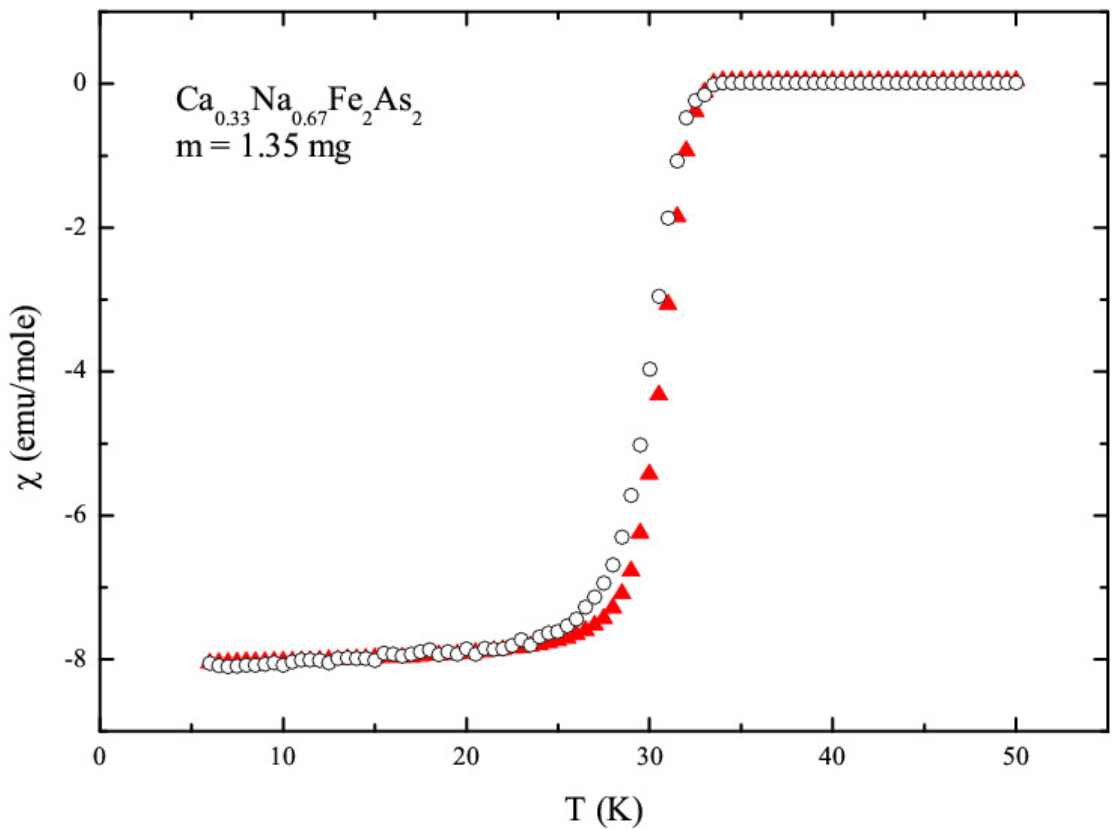

Fig. 1 (color online). Zero field cooled magnetic susceptibility of a single crystal of Ca$_{0.33}$Na$_{0.67}$Fe$_2$As$_2$ in contact with (solid red triangles) and without (open circles) Wakefield grease.

## III. Results and Discussion

The specific heat of a collage of 9.83 mg of single crystals (attached to the sample platform using Wakefield grease) of $Ca_{0.33}Na_{0.67}Fe_2As_2$ from 0.4 to 40 K is shown in Fig. 2. The measured value of $\Delta C/T_c = 105 \pm 5$ mJ/molK$^2$ is plotted in the revised [9] BNC plot shown in Fig. 3. Clearly, the current results exceed

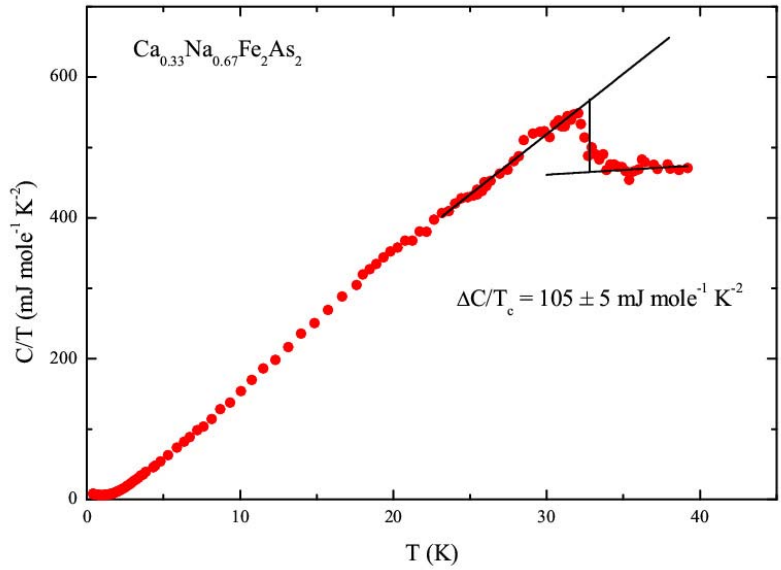

Fig. 2 (color online) Idealized transition at $T_c$ for the discontinuity in the specific heat, $\Delta C$, divided by $T_c$ for $Ca_{0.33}Na_{0.67}Fe_2As_2$ the trend established by BNC [8], for all the measured IBSs, and are consistent with the value (100 mJ/molK$^2$) for $\Delta C/T_c$ for single crystal $Ba_{0.6}K_{0.4}Fe_2As_2$, $T_c = 34.6$ K.

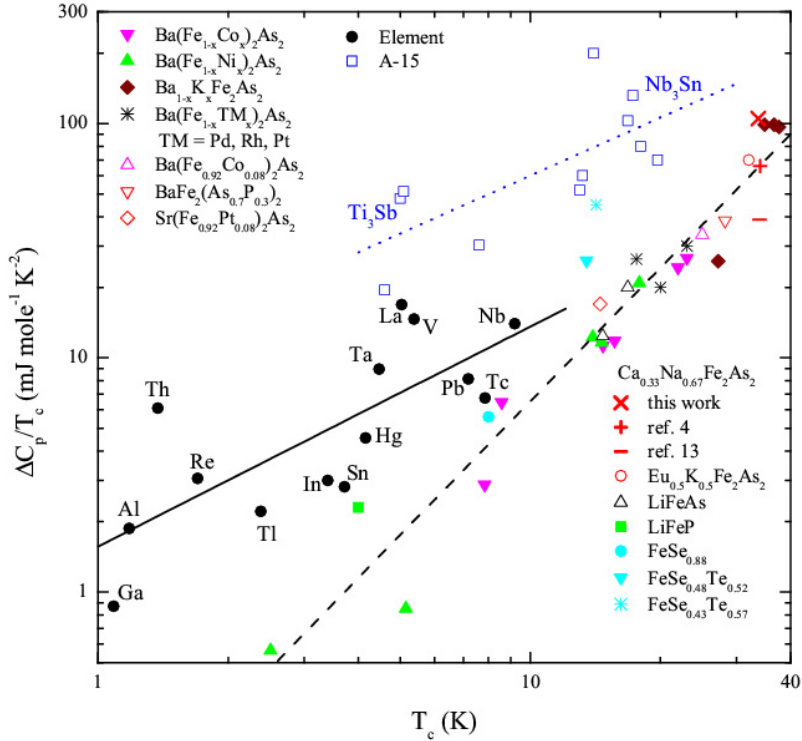

Fig. 3 (color online) BNC [8] plot, as revised in ref. 9, of $\Delta C/T_c$ vs $T_c$ values for both the iron based superconductors, the superconducting elements with $T_c>1$ K, and a selection of A-15 structure superconductors. As may be seen, such a plot seems to indicate a different behavior (see refs. 2 and 9 for a discussion) for the iron based superconductors vs the conventional superconductors. The data for $Ca_{0.33}Na_{0.67}Fe_2As_2$ from the present work, and refs. 4 and 13, show a huge spread in magnitude (from 39 [13] to 105 mJ/molK$^2$ [present work]), indicating the need for attention to sample quality *and* the utility of the BNC plot as a metric for this.

Thus, there is no question of inhomogeneous superconductivity or multi-phase sample for the current work on $Ca_{0.33}Na_{0.67}Fe_2As_2$. The bulk of this sample is a superconductor, which is consistent with the rather low extrapolation of C/T to T=0, $\gamma \approx 0.8$ mJ/molK$^2$, in the superconducting state shown below in Fig. 4.

Numerous attempts for IBSs exist in the literature (for an early example [12] see the work on $Ba(Fe_{0.93}Co_{0.07})_2As_2$) to apply a multi-band fitting model to such zero field specific heat data as shown in Fig. 2 in order to gain insight into the band gaps, pairing symmetries, and coupling strengths. We believe that the number of fitting parameters involved in such multi-band models (a recent specific heat paper [13] on $Ca_{0.32}Na_{0.68}Fe_2As_2$ uses a three band model and discusses extension to four and five band models) make the value of such fits questionable. In particular, if it is not known if the sample specific heat is characteristic of ideal, intrinsic behavior for the compound then such analyses are futile. Although the measured value of $\Delta C/T_c = 105 \pm 5$ mJ/molK$^2$ from the data in Fig. 2, and plotted in the revised BNC plot from ref. 9 in Fig. 3, for our sample of $Ca_{0.33}Na_{0.67}Fe_2As_2$ seems indicative of 'good quality', sample quality - as in *all* IBS - remains an issue. For example, the recent work [13] on single crystals of $Ca_{0.32}Na_{0.68}Fe_2As_2$ with a comparable $T_c^{bulk}$ onset value of $\approx$34 K reports only $\Delta C/T_c = 39$ mJ/molK$^2$ (see the point plotted in the revised BNC plot in Fig. 3), as well as a 'significant hump' in C/T at 13 K which is absent from our data in Fig. 2.

Refinement of sample quality is still ongoing in $Ca_{1-x}Na_xFe_2As_2$, with the possibility of an even larger value of $\Delta C/T_c$ than that reported here. As another metric of sample quality, we present the zero field specific heat of $Ca_{0.33}Na_{0.67}Fe_2As_2$ from 0.4 to 3.5 K in Fig. 4. The upturn in the data plotted as C/T as temperature is lowered below approximately 1 K is thought to be due to a magnetic second phase, a common occurrence [14] in the specific heat of IBSs.

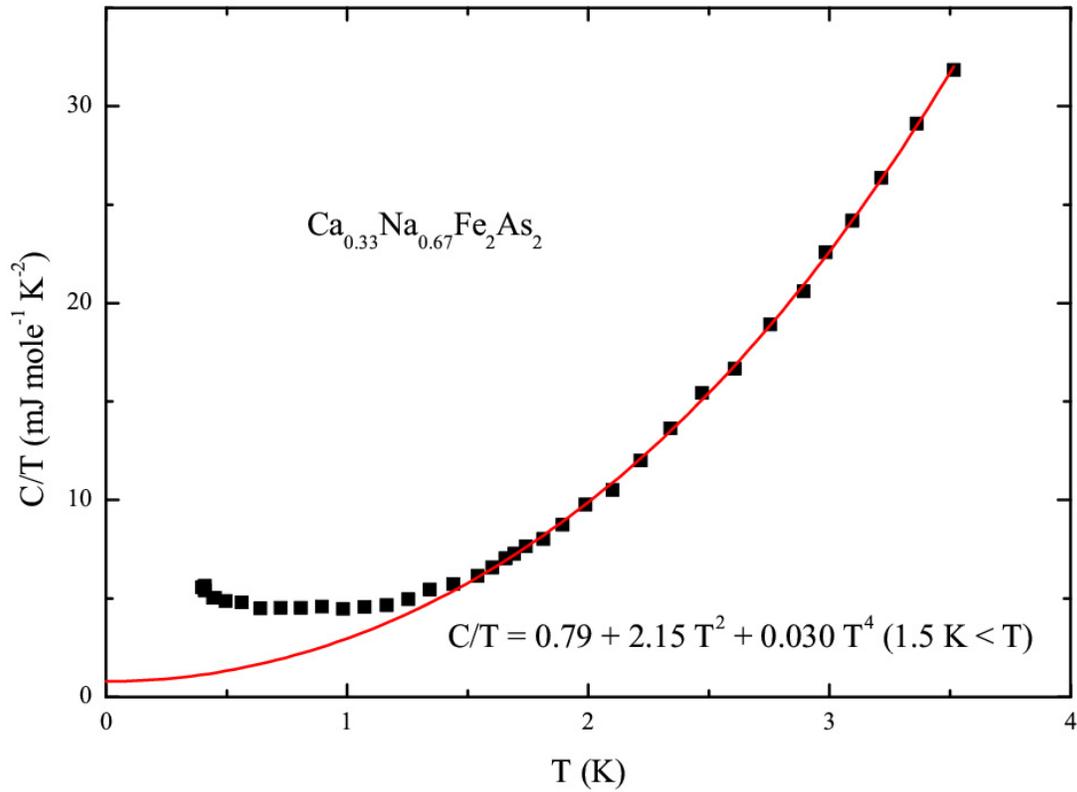

Fig. 4 (color online) Specific heat divided by temperature, C/T, vs temperature for $T \leq 3.5$ K for $Ca_{0.33}Na_{0.67}Fe_2As_2$. Note the low temperature upturn in C/T below 1.5 K. The extrapolation of these zero field C/T data to T=0 results in a residual $\gamma$ value of 0.79 mJ/molK$^2$.

IV. Conclusions

$Ca_{0.33}Na_{0.67}Fe_2As_2$ is clearly a bulk superconductor, with a low value ($\approx 0.8$ mJ/molK$^2$) of C/T extrapolated to T=0 in the superconducting state and a value of $\Delta C/T_c$ which exceeds by 50% the BNC 'average' trend line [8] where $\Delta C/T_c$ scales approximately as $T_c^2$ for all the IBSs. Perhaps, as further progress is made on improving sample quality, the five year old original [8] fitted line through the BNC $\Delta C/T_c$ vs $T_c$ data for iron based superconductors will need to be raised. Recent careful work [15] on polycrystalline samples of $Ba_{1-x}Na_xFe_2As_2$ with $T_c$ values approaching 34 K shows values of $\Delta C/T_c$ even 10% larger than the value reported here for single crystal $Ca_{0.33}Na_{0.67}Fe_2As_2$.

Acknowledgements: Work at Florida performed under the auspices of the US Department of Energy, Basic Energy Sciences, contract no. DE-FG02-86ER45268. Work in China performed is supported by NSF & MOST of China through research Projects.